\documentclass[aps,pra,twocolumn,superscriptaddress,eqsecnum]{revtex4}
\usepackage{amsfonts}%
\usepackage{amsmath}%
\usepackage{amssymb}%
\usepackage{color,graphicx}

\newcommand{\beq}{\begin{equation}}
\newcommand{\eeq}{\end{equation}}
\newcommand{\bqa}{\begin{eqnarray}}
\newcommand{\eqa}{\end{eqnarray}}

\newcommand{\erf}[1]{Eq.~(\ref{#1})}

\newcommand{\dg}{^\dagger}
\newcommand{\rt}[1]{\sqrt{#1}\,}

\newcommand{\bra}[1]{\langle{#1}|}
\newcommand{\ket}[1]{|{#1}\rangle}

\newcommand{\cu}[1]{\left\{ {#1} \right\}}
\newcommand{\ro}[1]{\left( {#1} \right)}

\definecolor{nblue}{rgb}{0.2,0.2,0.7}
\definecolor{ngreen}{rgb}{0.2,0.6,0.2}
\definecolor{nred}{rgb}{0.7,0.2,0.2}
\definecolor{nyellow}{rgb}{0.7,0.6,0.2}
\definecolor{npurple}{rgb}{0.7,0.1,0.7}
\definecolor{nbackground}{rgb}{1,1,1}
\definecolor{ngrey}{rgb}{0.5,0.5,0.5}
\definecolor{nbrown}{rgb}{0.6,0.4,0.2}
\definecolor{nblack}{rgb}{0,0,0}

\begin{document}
\title{Pooling quantum states obtained by indirect measurements}
\author{Robert~W.~Spekkens}
\affiliation{Department of Applied Mathematics and Theoretical
Physics, University of Cambridge, Cambridge CB3 0WA, United Kingdom}
\author{H. M.
Wiseman} \affiliation{Centre for Quantum Computer Technology, Centre
for Quantum Dynamics, School of Science, Griffith University,
Brisbane, 4111 Australia}
\date{Dec. 20, 2006}

\begin{abstract}
We consider the pooling of quantum states when Alice and Bob  both have
  one part of a tripartite system and, on the basis of
measurements on their respective  parts, each infers  a quantum state
for the third  part  $S$. We denote the conditioned states which
Alice and Bob assign to $S$ by $\alpha$ and $\beta$ respectively,
while the unconditioned state of $S$ is $\rho$. The state assigned
by an overseer, who has all the data available to Alice and Bob, is
$\omega$. The pooler is told only $\alpha$, $\beta$, and $\rho$.  We
show that for certain classes of tripartite states, this information is
enough for her to reconstruct $\omega$ by the formula $\omega
\propto \alpha \rho^{-1} \beta$. Specifically, we identify two classes
of states for which this pooling formula works:
(i) all pure states for which the rank of $\rho$ is equal to the product of
the ranks of the states of Alice's and Bob's subsystems;
(ii) all mixtures of tripartite product states that are mutually orthogonal on $S$.
\end{abstract}

\maketitle

\section{Introduction}

\strut In the approach to quantum theory wherein a quantum state
encodes an observer's knowledge of a system, different observers may
assign different quantum states. \ It is then natural to ask which
pairs of quantum states are compatible in the sense that they can be
simultaneously assigned by a pair of observers \cite{BFM02,Bru02}.
Another natural question concerns pooling
\cite{Bru02,Jac02,PB03,Her04,Jac05,Bru06}: what is the quantum state
that ought to be assigned to a system by someone who learns only the
quantum states assigned to it by two distinct observers?

The way in which we have just posed the pooling problem --- the
standard way to do so --- presumes that no further assumptions need
to be specified in order for there to be a unique solution. That
this is not necessarily the case becomes clear when we examine
carefully the notion of pooling states of knowledge in classical
probability theory. Such an analysis is particularly appropriate
given that the problem of pooling quantum states was originally
conceived as an analogue of classical pooling.

The classical pooling result may be stated in the following way.
Suppose that Alice and Bob share a prior probability distribution
$p(s)$ over a space of hypotheses.  Suppose further that the data
they acquire before updating their distributions is obtained
\emph{independently}. Specifically, if $a$ denotes Alice's data and
$b$ denotes Bob's data, then these are assumed to be conditionally
independent given $s$: \beq p(a,b|s) =p(a|s)p(b|s).
\label{condind}\eeq In these circumstances, Alice, upon learning
$a$, updates her description from $p(s)$ to $p(s|a) \propto p(a|s)p(s)$
and Bob, upon learning $b$, updates his description from $p(s)$ to
$p(s|b) \propto p(b|s)p(s)$.
A third party who has access to both data $a$ and $b$, whom we shall
call the overseer and name Oswald, would update his description from
$p(s)$ to $p(s|a,b)$, where
\begin{equation}
\label{classicalpooling}
 p(s|a,b) \propto \frac{p(s|a)p(s|b)}{p(s)}\;,
 \end{equation}
as one easily verifies by applying Bayes' theorem to \erf{condind}.
In fact, \erf{classicalpooling} holds only for those values of $s$
for which $p(s)> 0$; for values of $s$ for which $p(s)=0$, Oswald
assigns $p(s|a,b)=0$. The important point to note is that the
overseer's probability distribution depends only on the prior
distribution and Alice's and Bob's posterior distributions.  Thus,
another party, who only has knowledge of these three distributions,
can reconstruct Oswald's distribution. This person, whom we shall
call the pooler and name Penelope,  need not  know any
additional details of how Alice and Bob came to update the prior in
the way that they did.

However, the formula (\ref{classicalpooling}) does not hold in
general, as discussed in  Sec.~\ref{othercases}.
In such cases, Oswald, who is aware of how the data was acquired,
can still use Bayes' theorem to update his distribution. But
Penelope, who only has the prior distribution and Alice's and Bob's
posterior distributions at her disposal,  has insufficient
information to reconstruct Oswald's state of knowledge.  It is only
under special circumstances, such as when \erf{condind} holds,
that she is able to do so.

The lessons of classical pooling for quantum pooling are several.
First, we have seen that in the one case where the pooler can
reconstruct the overseer's posterior distribution,  she must
use  not only Alice and Bob's distributions, but also the prior
distribution. Consequently, we expect that in the quantum case
she will require not only Alice's and Bob's quantum states, but also
the ``prior'' quantum state --- the
one that they both assigned prior to the measurements. Second, in
order to tackle the pooling problem, it has been useful to imagine
an overseer who knows every aspect of the protocol and the collected
data, because his posterior is guaranteed to be uniquely specified
by Bayes' theorem. This suggests that in the quantum case, we again ought
to consider an overseer  of this nature, because the quantum state that the overseer
must assign after the measurements --- the ``posterior'' quantum
state --- will be uniquely specified by quantum theory (through the
quantum update rule).  If this state is a function only of Alice's state,
Bob's state, and the prior state, then Penelope can succeed in her task
of reconstruct Oswald's
posterior state. Third, we expect that there are limited
circumstances in which the pooler can succeed in this fashion.

Most of the previous work on pooling has not taken this approach.
Typically, the prior quantum state is not specified in the problem,
nor is it specified how Alice and Bob acquired their
data~\cite{Bru02,PB03,Her04}; the notion of pooling that is
discussed in these papers is therefore distinct from the one
considered here. On the other hand, Jacobs \cite{Jac02,Jac05} and
Brun in Ref.~\cite{Bru06} have emphasized the importance of the
method of data acquisition for pooling knowledge about a system,
have taken care to specify a prior quantum state (they both assume a
completely mixed state), and have used the device of an overseer
 to evaluate pooling
strategies. The pooling problem that Jacobs considers is nonetheless
distinct from the one considered here. This is because the two
states to be pooled in his approach are not simultaneous
descriptions of a single system. Rather, they are descriptions of a
single system at two times, between which there is an intervening
direct measurement on the system, a measurement which may well
invalidate the applicability of the earlier description \footnote{It
follows that Alice's and Bob's quantum states need not be
compatible, rendering the constraints on pooling that one derives
from the compatibility constraint \cite{Jac02,Her04} (discussed in
Sec.~\ref{subsec:compatibility}) inapplicable in this context, a
fact which Jacobs acknowledges.}. However, a particular instance of
state pooling considered by Brun in Ref.~\cite{Bru06} (wherein Alice
and Bob obtain conditionally independent data about the outcome of a
measurement on the system) \emph{does} fall within the general framework
 described above.

\begin{figure}[ht]
\begin{center}
\includegraphics[width=85mm]{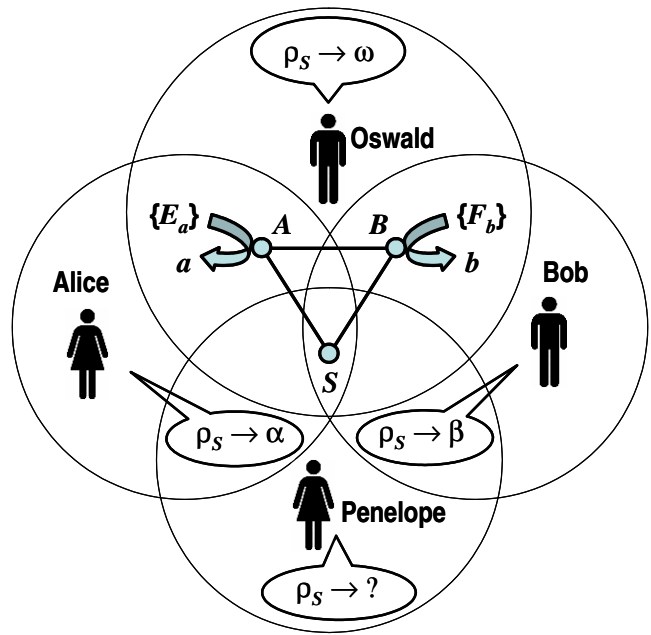}
\caption{Schematic of the pooling problem we consider. All parties
seek to update the quantum state they assign to system $S$ based on
the information they acquire.  Each party has access to only certain
information, represented by the contents of the circle in which he
or she is enclosed. All information about $S$ is ultimately derived
from indirect measurements on $S$, that is, measurements upon
systems ($A$ and $B$) that are correlated with $S$. The prior
knowledge about $A$, $B$, and $S$ of the different parties derives
from a single tripartite state, but each party has the reduced state
for only those systems contained in his or her circle. Alice
performs a measurement on $A$ (represented by a POVM $\{E_a\}$) with
outcome $a$, and updates her state for $S$ to $\alpha$. Bob does
likewise, {\em mutatis mutandis}.  Oswald, the overseer, learns
everything known to Alice and Bob and updates his state to $\omega$.
Penelope, the pooler, learns only the manner in which Alice and Bob
update their states. The question we address is: what state should
she assign to $S$? } \label{fig1}
\end{center}
\end{figure}

In the present article, we consider the  specific case  where Alice's and Bob's
data are obtained by measurements upon two different shares of a
tripartite system prepared in a quantum state known to them both.
Based upon this data, they calculate updated states for the third
share. We identify two classes of tripartite states for which
Alice's and Bob's updated states,
together with their initial state (for this third share), enable
Penelope to reconstruct Oswald's state, in analogy to
\erf{classicalpooling}. This is illustrated in Fig.~\ref{fig1}. We
also show that our formula fails in more general cases, as one would
expect. We discuss the relation of our results to state
compatibility, and generalize our results to the case of arbitrarily
many parties.

\section{Pooling quantum states from indirect measurements} \label{sec2}

The scenario we consider is as follows \footnote{It has previously
been considered by Brun, Finkelstein and Mermin~\cite{BFM02} in the
context of compatibility and by Jacobs~\cite{Jac02} in the context
of pooling.}. There is a tripartite system $ABS$ for which Alice
possesses the $A$ share and Bob possesses the $B$ share. They both
initially describe the tripartite system by the (possibly mixed)
quantum state $W$ defined on $\mathcal{H}^{A}\otimes\mathcal{H}^{B}
\otimes\mathcal{H}^{S},$
 and thus they both initially describe the third system $S$ by the same
reduced state $\rho\equiv\text{Tr}_{AB}\left(W\right)$.
 Alice makes a generalized measurement on $A,$
 and Bob makes a generalized measurement on $B.$ After registering the
outcome of her measurement, Alice updates her description of the
system $S$ to a new quantum state which we denote by $\alpha$.
Likewise, after registering the outcome of his measurement, Bob
updates his description of $S$ to a new quantum state which we
denote by $\beta.$ Quantum theory uniquely specifies how Alice and
Bob ought to update their descriptions.  (Strictly speaking, Alice
only requires the reduction of $W$ to $AS$ and Bob only requires the
reduction to $BS$, as is indicated schematically in
Fig.~\ref{fig1}.)

The overseer, Oswald, knows everything there is to know: the initial
state $W,$  what measurements Alice and Bob made, and the outcomes
they obtained. After learning all of this information, the overseer
updates his description of $S$ to the state $\omega.$ Again, quantum
theory uniquely specifies how the overseer ought to update his
description.

The pooler, Penelope, is not given any information about what
measurements are made by Alice and Bob, nor their outcomes. Rather,
she is told only the identity of the initial quantum state for $S$
as well as the quantum states that Alice and Bob assign to $S$ at
the end of their measurements. In other words, the pooler is given
only $\rho$, $\alpha$ and $\beta.$ Her task is to \textquotedblleft
pool\textquotedblright\ the information contained in $\alpha$ and
$\beta$ (and possibly also $\rho$) to obtain a single quantum state
that will represent what she predicts about the outcomes of future
experiments on $S.$

By analogy with the classical case, we expect that the quantum
pooling problem only has a solution when Alice's and Bob's data
satisfy some condition of independence.  Although we do not
determine the most general condition here, we do illustrate two
simple classes of states $W$ for which the quantum pooling problem can be solved.
Class (i) is the class of pure tripartite states in which the rank of $\rho$
is maximal (in a sense defined below). Class (ii) is the class of
mixtures of product states that are mutually orthogonal on $S$.

We show that for both classes, Oswald's quantum state $\omega$
can be written as a simple function of $\alpha,$ $\beta$, and
$\rho$, namely, \bqa
\omega & \propto &\alpha\rho^{-1}\beta \label{beauty} \\
& = &\beta\rho^{-1}\alpha, \label{bottom} \eqa where $\rho^{-1}$ is
the inverse of $\rho$ on its support \footnote{The support of an
operator $A$ is the span of the eigenspaces of $A$ having non-zero
eigenvalue; it is the orthogonal complement of the kernel of $A$.
The inverse of $A$ on its support is the operator obtained by
replacing the non-zero eigenvalues of $A$ by their inverses.}. To
obtain the precise form of $\omega$, one need only normalize the
right-hand side of this expression. Penelope can obviously construct
this state, and so for these classes, she succeeds in the pooling task.

\subsection{Class (i)} \label{case1}

Let $W\equiv \ket{\Psi}\bra{\Psi}$, where $\ket{\Psi}$ is a tripartite
state. We define $\mathcal{H}^{X}$ ($X=A,B,S$) to be the support
of $W$ on subsystem $X$. That is, if
$d_X$ is the rank of the reduced state of the subsystem $X$,
this is also the dimensionality of $\mathcal{H}^X$.
Now restrict to the class of pure states where $d_S = d_A d_B$. Note that
this is the {\em maximal} value $d_S$ can take, given $d_A$ and $d_B$, because
(from the purity of $W$) $d_S = d_{AB} \leq d_A d_B$. By the
Choi-Jamiolkowski isomorphism \cite{Jam72},  any $\ket{\Psi}$
satisfying this condition  can be
written as
\begin{equation} \label{ChoiJam}
\left\vert \Psi\right\rangle =\sum_{e=1}^{d_A}\sum_{f=1}^{d_B}
\left\vert e,f\right\rangle \otimes(\rt{\rho}U\left\vert e,f\right\rangle). %
\end{equation}
Here $\{\left\vert e,f\right\rangle \}$ is a complete orthonormal
basis for $\mathcal{H}^{A}\otimes \mathcal{H}^{B}$, and $U$ is a unitary
operator from
 $\mathcal{H}^{A}\otimes \mathcal{H}^{B}$ to ${\cal H}^S$, so that
  $U\dg U = I^A\otimes I^B$.

Alice's generalized measurement can be represented by a
positive-operator-valued-measure  (POVM) $\cu{E_a}$. If she
gets result $a$ then the relevant POVM element (also called an {\em
effect})
 is the positive operator
$
E_a\otimes I \text{ on }\mathcal{H}^{A}\otimes\mathcal{H}^{B}.
$
She updates her description of $S$ from $\rho$ to
\begin{align}
\label{alpha}
\alpha &\propto {\rm Tr}_{AB}[(E_a\otimes I)\ket{\Psi}\bra{\Psi}] \nonumber \\
&= \sum_{e,f,e',f'} \langle e',f' \vert E_a\otimes I \vert e,f
\rangle \rt{\rho}U\left\vert e,f\right\rangle\langle
e',f' \vert U^{\dag} \rt{\rho} \nonumber \\
&= \rt{\rho} U(E_a\otimes I)^{T}U\dg \sqrt{\rho}\,,
\end{align}
where  $M^{T}$ denotes the transpose of $M$  in the basis $\ket{e,f}$ and
where we have made use of the completeness of this basis. Note that
the update rule for indirect measurements
has the POVM element sandwiched between square roots of the state.
This is the opposite of what occurs with direct measurements where
the state is typically sandwiched between square roots of the POVM
element.  Equation (\ref{alpha}) is an instance of the
Hughston-Jozsa-Wootters theorem~\cite{HJW93} generalized  to POVM
measurements~\cite{Fuc02,SR02}. The trace of the right-hand-side of
Eq.~(\ref{alpha}) is the probability for Alice to have obtained the
result she did, and $\alpha$ is simply the right-hand-side divided
by this quantity.

Similarly, if the POVM that Bob measures is $\cu{F_b}$, the effect
corresponding to the result $b$ is
$
I\otimes F_b\text{ on }\mathcal{H}^{A}\otimes\mathcal{H}^{B}.
$
Then he updates his description of $S$ from $\rho$ to
\begin{equation}
\label{beta}
 \beta \propto \rt{\rho}U(I\otimes F_b)  ^{T}
U\dg \sqrt{\rho}.
\end{equation}
The overseer, upon being told Alice's and Bob's outcomes, updates
his description of $S$ from $\rho$ to
\begin{equation}
\label{omega}
 \omega\propto \rt{\rho}U(E_a\otimes F_b)  ^{T}
U\dg \sqrt{\rho}.
\end{equation}

It follows from the above results that \
\begin{align}
\omega & \propto \rt{\rho} U (I\otimes F_b)^T(E_a\otimes
I)^T U^{\dag}\sqrt{\rho} \nonumber \\
& =\rt{\rho} U(I\otimes F_b)^T U\dg U(E_a\otimes I)^T U\dg \sqrt{\rho} \nonumber \\
& =\rt{\rho}U(E_a\otimes I)^T U\dg\rt{\rho}\rho^{-1}
\rt{\rho}U(I\otimes F_b)^{T}U\dg\sqrt{\rho} \nonumber \\
& \propto \alpha\rho^{-1}\beta. \label{finalomega}
\end{align}
Here we have made use of the fact that $U\dg U$ is the identity operator
on $\mathcal{H}^{A}\otimes\mathcal{H}^{B},$ the fact that $(AB)^{T}=B^{T}%
A^{T},$ and the fact that $\sqrt{\rho}\rho^{-1}\sqrt{\rho}$ is the
identity operator on $\mathcal{H}^{S}$ (by virtue of the rank condition on $\rho$).


Thus we have shown that Penelope should pool the states according to
\erf{beauty}. The proof of the second identity (\ref{bottom}) is
identical, except that the order of $(I\otimes F_b)$ and
$(E_a\otimes I)$ (which commute) is inverted in the first line of
\erf{finalomega}.

\subsubsection{Example}

To aid the reader's understanding, we now give an explicit application of the
above case. Consider the following state in $\mathcal{H}^A\otimes\mathcal{H}^B\otimes\mathcal{H}^S$:
\beq \label{224state}
\ket{\Psi} \propto \ket{0}\ket{0}\ket{0}+\ket{0}\ket{1}\ket{1}+\ket{1}\ket{0}\ket{2} + \ket{1}\ket{1}\ket{3},
\eeq
which satisfies the rank condition for class (i) states since $d_S=4$ and $d_A=d_B=2$.
It is obvious that if Alice and Bob measure in the logical
($\ket{0}, \ket{1}$) basis then the pooling formula will hold, as
each of them holds  independent classical information about the logical state of $S$.
It is less obvious that \erf{beauty} holds regardless of the measurements they  make.

Suppose, for example, that Alice and Bob both measure in the
($\ket{+}, \ket{-}$) basis, where $\ket{\pm} \propto \ket{0}\pm\ket{1}$. Say they both obtain the result $+$. Then it is easy to verify that Alice's updated state for $S$ is
\beq
\alpha \propto \uplus \ro{ \ket{0}+\ket{2} } \uplus \ro{\ket{1}+\ket{3}}.
\eeq
Here, following Ref.~\cite{Jon06}, we are using the following notation,
that for an arbitrary {\em ray}
$\ket{r}$, we have $\uplus \ket{r} \equiv + \ket{r}\bra{r}$.
Similarly, Bob's new state assignment is
\beq
\beta \propto \uplus \ro{ \ket{0}+\ket{1} } \uplus \ro{\ket{2}+\ket{3}},
\eeq
while Oswald's state (which is pure) is
\beq
\omega \propto \uplus \ro{ \ket{0}+\ket{1} + \ket{2}+\ket{3}} .
\eeq
Now for the state (\ref{224state}), $\rho \propto I$, from which
 it is easy to verify that \erf{beauty} holds.

\subsubsection{Pure states for which the pooling formula fails} \label{counter}

If a pure state does not satisfy
the rank condition for class (i) states, then in general the pooling formula does not hold.
This can be seen from the Greenberger-Horne-Zeilinger state
\beq \label{222state}
\ket{\Psi} = \ket{0}\ket{0}\ket{0}+\ket{1}\ket{1}\ket{1}.
\eeq
Here $d_S = d_A = d_B = 2$ so that $d_S \neq d_Ad_B$.
As with state (\ref{224state}), if Alice and Bob both measure 
in the logical basis
then the pooling formula will hold.
But in contrast with state (\ref{224state}),
this is no longer true if they measure
 in other bases.

Suppose, as above, that Alice and Bob both measure in the
($\ket{+}, \ket{-}$) basis, and both obtain the result $+$. Then it is easy to verify that Alice
and Bob obtain no information about $S$:
\beq
\alpha = \beta = \rho \propto  \uplus \ket{0} \uplus \ket{1} = I.
\eeq
By contrast, Oswald's updated state assignment is pure:
\beq
\omega \propto \uplus \ro{ \ket{0} + \ket{1} }.
\eeq
Clearly \erf{beauty} fails.

\subsection{Class (ii)} \label{case2}

Class (ii) states are separable mixed states of the
form \beq \label{mixedpoolok}
W=\sum_s p(s)\, (\sigma_s \otimes \tau_s \otimes \rho_s), \eeq
where $\sigma_s$,
$\tau_s$ and $\rho_s$ are normalized (possibly mixed) quantum states defined on
$\mathcal{H}^A$, $\mathcal{H}^B$ and $\mathcal{H}^S$ respectively.
We also require that the the different $\rho_s$ be defined on different subspaces:
\beq
\rho_s \rho_{s'} = 0 \textrm{ if } s\neq s'.
\eeq
 It follows that the initial state of system $S$ is
\beq \rho = \sum_s p(s)\, \rho_s\,. \eeq When Alice's measurement
reveals the outcome associated with the effect $E_a\otimes I$, she
updates her description of $S$ to \bqa \alpha &=& \sum_s p(s|a)
\rho_s\,, \label{case2alpha}\eqa where $ p(s|a) \propto  p(s){\rm
Tr}_A(E_a \sigma_s)$. Similarly, when Bob's measurement reveals the
outcome associated with the effect $I \otimes F_b$, he updates his
description of $S$ to \bqa \beta &=& \sum_s p(s|b) \rho_s\,,
\label{case2beta}\eqa where $p(s|b)\propto p(s){\rm Tr}_B(F_b
\tau_s)$. Oswald, upon learning $a$ and $b$, updates his description
to \bqa \omega &=& \sum_s p(s|a,b) \rho_s\,, \label{case2omega}\eqa
where $p(s|a,b)\propto p(s){\rm Tr}_A(E_a \sigma_s){\rm Tr}_B(F_b
\tau_s)$. Noting that $p(s|a,b) \propto p(s|a)p(s|b)/p(s)$ and that
$\alpha$, $\beta$, and $\rho$ commute, it follows that Penelope can
construct $\omega$ from $\alpha$, $\beta$ and $\rho$ according to
\erf{beauty}.

\label{othercases}
In Sec.~\ref{counter} we gave a pure state example for which
the pooling formula (\ref{beauty}) fails. It is also simple to find a mixed
state example. It suffices to consider the
tripartite state $W=\sum_{a,b,s} \,q(a,b,s)\, \ket{a}\bra{a} \otimes
\ket{b}\bra{b} \otimes \ket{s}\bra{s}$, where $\ket{a},$ $\ket{b},$
and $\ket{s}$ form orthonormal bases for systems $A$, $B$ and $S$.
For the case of measurements $E_a=\ket{a}\bra{a}$ and
$F_b=\ket{b}\bra{b}$, it is clear that the problem becomes
essentially classical. Specifically, Eqs.~(\ref{case2alpha}),
(\ref{case2beta}) and (\ref{case2omega}) hold, but where $p(s),
p(s|a), p(s|b)$ and $p(s|a,b)$ are now defined from $q(a,b,s)$.
Given that $\alpha$, $\beta$, and $\rho$ commute, the condition for
the quantum pooling formula, $\omega \propto \alpha\rho^{-1}\beta$,
to hold is that the classical pooling formula, $p(s|a,b) \propto
p(s|a)p(s|b)/p(s)$, holds. However, as mentioned in the
introduction, there are joint probability distributions $q(a,b,s)$
for which the latter fails.   For example, suppose
$a$ is correlated imperfectly with $s$, but
$b$ is {\em perfectly} correlated with $a$. Then the information in
$a$ and $b$ is redundant: $p(s|a,b) = p(s|a) = p(s|b)$. But the pooling
formula multiplies $p(s|a)$ and $p(s|b)$, yielding a distribution that is
too narrow in general. Explicit instances using Gaussian distributions
are easy to construct.


%

\section{Discussion}

\subsection{The connection to compatibility}
\label{subsec:compatibility}

Two classical probability distributions are compatible if and only
if their supports on the space of hypotheses (the regions to which
they assign non-zero probability) have some overlap. The same
concept was applied to quantum states by Brun, Finkelstein and
Mermin (BFM) \cite{BFM02}: two quantum states are compatible if and
only if the intersection of their supports on the Hilbert space is
not null \footnote{Caves, Fuchs, and Schack, however, dispute the
uniqueness of this condition, as well as the one for classical
compatibility  \cite{CFS02}.}.  This condition is equivalent to the
requirement that one can find a quantum state that appears with
non-zero weight in some convex decomposition of $\alpha$ and also in
some convex decomposition of $\beta$. Specifically, there must exist
a state $\omega$ and nonzero weights $p$ and $q$ such that \beq
\alpha=p\omega + (1-p)\alpha' \;,\;\; \beta=q\omega + (1-q)\beta'
\label{compatibility} \eeq for some states $\alpha'$ and $\beta'$.

 It is easy to verify that a pair of quantum states $\alpha, \beta$ for $S$ that
are acquired by indirect measurements (as we have been considering)
are compatible according to the BFM criteria. In the two cases we
have considered above it is also easy to verify that Penelope's
state $\omega$ (which is Oswald's state) does lie in the
intersection of the supports of Alice's and Bob's states.
Consequently, $\omega$ is incompatible with all states that are
incompatible with either Alice's or Bob's states. This requirement
for pooling has been previously emphasized by Jacobs \cite{Jac02}
and by Herbut \cite{Her04}.

\subsection{Generalization to an arbitrary number of parties}

The above results of Sec.~\ref{sec2} are
straightforward to generalize to an arbitrary number $N$ of parties:
Alice, Bob,~\ldots,~Zane. In the context of classical probability
theory, if the parties' data  are  conditionally independent, \beq
p(a,b|s) =p(a|s)p(b|s) \cdots p(z|s)\, \eeq then \beq
p(s|a,b,\ldots,z) \propto \frac{p(s|a)p(s|b)\cdots
p(s|z)}{p(s)^{N-1}}\,. \eeq

The quantum pooling formula for $N$ parties, which is the
generalization of \erf{beauty}, is \beq  \omega = \alpha \rho^{-1}
\beta \rho^{-1}\cdots \zeta\,,  \eeq where $\zeta$ is
Zane's posterior quantum state.  By proofs analogous to those
presented for the 2-party case, this formula can be shown to apply
if the initial $N$-partite quantum state $W$ is pure  with
$d_S = d_A d_B \cdots d_Z$, or if it is of the form
$W=\sum_s p(s)\, (\alpha_s \otimes \beta_s \otimes \cdots
\zeta_s \otimes \rho_s)$,
with $\rho_s\rho_{s'}=0$ for $s\neq s'$ as above.

\section{Conclusion}

Assuming quantum states are states of knowledge, one should
sometimes be able to pool the quantum states of different observers.
We have argued that it is critical to determine whether, from the
prior quantum state and Alice's and Bob's posterior quantum states,
one can reconstruct the quantum state that would be assigned by an
overseer who knew all the details of the experiment and was given
all of the data.  If this is the case, then the pooled quantum state
is simply the state assigned by the overseer. We have considered the
scenario wherein Alice and Bob acquire their data by indirect
measurements on the system, specifically, by measurements upon
distinct ancillas. We have shown two forms of the initial
tripartite state for which the pooler can reconstruct the state of
the overseer.  Finally, we have demonstrated the connection to the
notion of compatibility and the generalization to multiple parties.

 To end,  we note that the classes of states we have identified
do not contain all states for which the pooling formula holds \cite{LS07}.
To generalize our results further, one could hope to obtain insight from the classical case.
As noted in the introduction, conditional independence of  Alice's and Bob's data, Eq.~(\ref{condind}), is a sufficient
condition for the applicability of the classical pooling formula,
Eq.~(\ref{classicalpooling}).  However, it is not a necessary condition, so
 in seeking the most general conditions
under which the quantum pooling formula holds, one is not simply
seeking a quantum analogue of the conditional independence of Alice's
and Bob's data. Further
investigations into these issues are underway \cite{LS07}.

\begin{acknowledgments}
The authors acknowledge stimulating discussions with Robin
Blume-Kohout, Todd Brun, Kurt Jacobs, Matt Leifer, and David Poulin.
We also acknowledge the organizers of the workshop ``Being Bayesian
in a Quantum World'', Konstanz, Germany, 2005, where a talk by Todd
Brun prompted the current investigation. Finally, we gratefully acknowledge an anonymous referee
for kindly pointing out the necessity of the rank condition for class (i).
R.W.S. receives support
from the Royal Society and from the European Union through the
Integrated Project QAP (IST-3-015848), SCALA (CT-015714), SECOQC,
and QIP IRC (GR/S821176/01). H.M.W. is supported by the ARC and the
State of Queensland.
\end{acknowledgments}

\end{document}